\documentstyle[12pt]{article}
\advance \topmargin by -\headheight
\advance \topmargin by -\headsep

\evensidemargin \oddsidemargin
\begin{document}

\topmargin -.6in

\def\rh{{\hat \rho}}
\def\alie{{\hat{\cal G}}}
\newcommand{\sect}[1]{\setcounter{equation}{0}\section{#1}}
\renewcommand{\theequation}{\thesection.\arabic{equation}}

\def\rf#1{(\ref{eq:#1})}
\def\lab#1{\label{eq:#1}}
\def\nonu{\nonumber}
\def\br{\begin{eqnarray}}
\def\er{\end{eqnarray}}
\def\be{\begin{equation}}
\def\ee{\end{equation}}
\def\eq{\!\!\!\! &=& \!\!\!\! }
\def\foot#1{\footnotemark\footnotetext{#1}}
\def\lb{\lbrack}
\def\rb{\rbrack}
\def\llangle{\left\langle}
\def\rrangle{\right\rangle}
\def\blangle{\Bigl\langle}
\def\brangle{\Bigr\rangle}
\def\llbrack{\left\lbrack}
\def\rrbrack{\right\rbrack}
\def\lcurl{\left\{}
\def\rcurl{\right\}}
\def\({\left(}
\def\){\right)}
\newcommand{\nit}{\noindent}
\newcommand{\ct}[1]{\cite{#1}}
\newcommand{\bi}[1]{\bibitem{#1}}
\def\lskip{\vskip\baselineskip\vskip-\parskip\noindent}
\relax

\def\tr{\mathop{\rm tr}}
\def\Tr{\mathop{\rm Tr}}
\def\v{\vert}
\def\bv{\bigm\vert}
\def\Bgv{\;\Bigg\vert}
\def\bgv{\bigg\vert}
\newcommand\partder[2]{{{\partial {#1}}\over{\partial {#2}}}}
\newcommand\funcder[2]{{{\delta {#1}}\over{\delta {#2}}}}
\newcommand\Bil[2]{\Bigl\langle {#1} \Bigg\vert {#2} \Bigr\rangle}  
\newcommand\bil[2]{\left\langle {#1} \bigg\vert {#2} \right\rangle} 
\newcommand\me[2]{\left\langle {#1}\bv {#2} \right\rangle} 
\newcommand\sbr[2]{\left\lbrack\,{#1}\, ,\,{#2}\,\right\rbrack}
\newcommand\pbr[2]{\{\,{#1}\, ,\,{#2}\,\}}
\newcommand\pbbr[2]{\lcurl\,{#1}\, ,\,{#2}\,\rcurl}
%
\def\a{\alpha}
\def\at{{\tilde A}^R}
\def\atc{{\tilde {\cal A}}^R}
\def\atcm#1{{\tilde {\cal A}}^{(R,#1)}}
\def\b{\beta}
\def\dc{{\cal D}}
\def\d{\delta}
\def\D{\Delta}
\def\eps{\epsilon}
\def\vareps{\varepsilon}
\def\g{\gamma}
\def\G{\Gamma}
\def\grad{\nabla}
\def\h{{1\over 2}}
\def\l{\lambda}
\def\L{\Lambda}
\def\m{\mu}
\def\n{\nu}
\def\o{\over}
\def\om{\omega}
\def\O{\Omega}
\def\p{\phi}
\def\P{\Phi}
\def\pa{\partial}
\def\pr{\prime}
\def\pt{{\tilde \Phi}}
\def\ra{\rightarrow}
\def\s{\sigma}
\def\S{\Sigma}
\def\t{\tau}
\def\th{\theta}
\def\Th{\Theta}
\def\ti{\tilde}
\def\wti{\widetilde}
\def\jc{J^C}
\def\bj{{\bar J}}
\def\sj{{\jmath}}
\def\bsj{{\bar \jmath}}
\def\bp{{\bar \p}}
\def\vp{\varphi}
\def\vt{{\tilde \varphi}}
\def\faa{Fa\'a di Bruno~}
\def\ca{{\cal A}}
\def\cb{{\cal B}}
\def\ce{{\cal E}}
\newcommand\sumi[1]{\sum_{#1}^{\infty}}   
\newcommand\fourmat[4]{\left(\begin{array}{cc}  
{#1} & {#2} \\ {#3} & {#4} \end{array} \right)}

%
\def\lie{{\cal G}}
\def\dlie{{\cal G}^{\ast}}
\def\elie{{\widetilde \lie}}
\def\edlie{{\elie}^{\ast}}
\def\hlie{{\cal H}}
\def\flie{{\cal F}}
\def\wlie{{\widetilde \lie}}
\def\f#1#2#3 {f^{#1#2}_{#3}}
\def\winf{{\sf w_\infty}}
\def\win1{{\sf w_{1+\infty}}}
\def\hwinf{{\sf {\hat w}_{\infty}}}
\def\Winf{{\sf W_\infty}}
\def\Win1{{\sf W_{1+\infty}}}
\def\hWinf{{\sf {\hat W}_{\infty}}}
\def\Rm#1#2{r(\vec{#1},\vec{#2})}          
\def\OR#1{{\cal O}(R_{#1})}           
\def\ORti{{\cal O}({\widetilde R})}           
\def\AdR#1{Ad_{R_{#1}}}              
\def\dAdR#1{Ad_{R_{#1}^{\ast}}}      
\def\adR#1{ad_{R_{#1}^{\ast}}}       
\def\KP{${\rm \, KP\,}$}                 
\def\KPl{${\rm \,KP}_{\ell}\,$}         
\def\KPo{${\rm \,KP}_{\ell = 0}\,$}         
\def\mKPa{${\rm \,KP}_{\ell = 1}\,$}    
\def\mKPb{${\rm \,KP}_{\ell = 2}\,$}    
%
\def\rlx{\relax\leavevmode}
\def\inbar{\vrule height1.5ex width.4pt depth0pt}
\def\IZ{\rlx\hbox{\sf Z\kern-.4em Z}}
\def\IR{\rlx\hbox{\rm I\kern-.18em R}}
\def\IC{\rlx\hbox{\,$\inbar\kern-.3em{\rm C}$}}
\def\IN{\rlx\hbox{\rm I\kern-.18em N}}
\def\IO{\rlx\hbox{\,$\inbar\kern-.3em{\rm O}$}}
\def\IP{\rlx\hbox{\rm I\kern-.18em P}}
\def\IQ{\rlx\hbox{\,$\inbar\kern-.3em{\rm Q}$}}
\def\IF{\rlx\hbox{\rm I\kern-.18em F}}
\def\IG{\rlx\hbox{\,$\inbar\kern-.3em{\rm G}$}}
\def\IH{\rlx\hbox{\rm I\kern-.18em H}}
\def\II{\rlx\hbox{\rm I\kern-.18em I}}
\def\IK{\rlx\hbox{\rm I\kern-.18em K}}
\def\IL{\rlx\hbox{\rm I\kern-.18em L}}
\def\one{\hbox{{1}\kern-.25em\hbox{l}}}
\def\0#1{\relax\ifmmode\mathaccent"7017{#1}%
B        \else\accent23#1\relax\fi}
\def\omz{\0 \omega}
%
\def\ltimes{\mathrel{\vrule height1ex}\joinrel\mathrel\times}
\def\rtimes{\mathrel\times\joinrel\mathrel{\vrule height1ex}}
%
\def\mark{\noindent{\bf Remark.}\quad}
\def\prop{\noindent{\bf Proposition.}\quad}
\def\theor{\noindent{\bf Theorem.}\quad}
\def\name{\noindent{\bf Definition.}\quad}
\def\exam{\noindent{\bf Example.}\quad}
\def\proof{\noindent{\bf Proof.}\quad}
%
%
\def\PRL#1#2#3{{\sl Phys. Rev. Lett.} {\bf#1} (#2) #3}
\def\NPB#1#2#3{{\sl Nucl. Phys.} {\bf B#1} (#2) #3}
\def\NPBFS#1#2#3#4{{\sl Nucl. Phys.} {\bf B#2} [FS#1] (#3) #4}
\def\CMP#1#2#3{{\sl Commun. Math. Phys.} {\bf #1} (#2) #3}
\def\PRD#1#2#3{{\sl Phys. Rev.} {\bf D#1} (#2) #3}
\def\PLA#1#2#3{{\sl Phys. Lett.} {\bf #1A} (#2) #3}
\def\PLB#1#2#3{{\sl Phys. Lett.} {\bf #1B} (#2) #3}
\def\JMP#1#2#3{{\sl J. Math. Phys.} {\bf #1} (#2) #3}
\def\PTP#1#2#3{{\sl Prog. Theor. Phys.} {\bf #1} (#2) #3}
\def\SPTP#1#2#3{{\sl Suppl. Prog. Theor. Phys.} {\bf #1} (#2) #3}
\def\AoP#1#2#3{{\sl Ann. of Phys.} {\bf #1} (#2) #3}
\def\PNAS#1#2#3{{\sl Proc. Natl. Acad. Sci. USA} {\bf #1} (#2) #3}
\def\RMP#1#2#3{{\sl Rev. Mod. Phys.} {\bf #1} (#2) #3}
\def\PR#1#2#3{{\sl Phys. Reports} {\bf #1} (#2) #3}
\def\AoM#1#2#3{{\sl Ann. of Math.} {\bf #1} (#2) #3}
\def\UMN#1#2#3{{\sl Usp. Mat. Nauk} {\bf #1} (#2) #3}
\def\FAP#1#2#3{{\sl Funkt. Anal. Prilozheniya} {\bf #1} (#2) #3}
\def\FAaIA#1#2#3{{\sl Functional Analysis and Its Application} {\bf #1} (#2)
#3}
\def\BAMS#1#2#3{{\sl Bull. Am. Math. Soc.} {\bf #1} (#2) #3}
\def\TAMS#1#2#3{{\sl Trans. Am. Math. Soc.} {\bf #1} (#2) #3}
\def\InvM#1#2#3{{\sl Invent. Math.} {\bf #1} (#2) #3}
\def\LMP#1#2#3{{\sl Letters in Math. Phys.} {\bf #1} (#2) #3}
\def\IJMPA#1#2#3{{\sl Int. J. Mod. Phys.} {\bf A#1} (#2) #3}
\def\AdM#1#2#3{{\sl Advances in Math.} {\bf #1} (#2) #3}
\def\RMaP#1#2#3{{\sl Reports on Math. Phys.} {\bf #1} (#2) #3}
\def\IJM#1#2#3{{\sl Ill. J. Math.} {\bf #1} (#2) #3}
\def\APP#1#2#3{{\sl Acta Phys. Polon.} {\bf #1} (#2) #3}
\def\TMP#1#2#3{{\sl Theor. Mat. Phys.} {\bf #1} (#2) #3}
\def\JPA#1#2#3{{\sl J. Physics} {\bf A#1} (#2) #3}
\def\JSM#1#2#3{{\sl J. Soviet Math.} {\bf #1} (#2) #3}
\def\MPLA#1#2#3{{\sl Mod. Phys. Lett.} {\bf A#1} (#2) #3}
\def\JETP#1#2#3{{\sl Sov. Phys. JETP} {\bf #1} (#2) #3}
\def\JETPL#1#2#3{{\sl  Sov. Phys. JETP Lett.} {\bf #1} (#2) #3}
\def\PHSA#1#2#3{{\sl Physica} {\bf A#1} (#2) #3}
\def\PHSD#1#2#3{{\sl Physica} {\bf D#1} (#2) #3}

\begin{titlepage}
\vspace*{-1cm}
\noindent
\\
\vskip .3in
\begin{center}
{\large\bf The Conserved Charges and Integrability of}
\end{center}
\begin{center}
{\large\bf  the Conformal Affine Toda Models}
\end{center}

\normalsize
\vskip .4in

\begin{center}
{ H. Aratyn\footnotemark
\footnotetext{Work supported in part by U.S. Department of Energy,
contract DE-FG02-84ER40173 and by NSF, grant no. INT-9015799}}

\par \vskip .1in \noindent
Department of Physics \\
University of Illinois at Chicago\\
845 W. Taylor St.\\
Chicago, Illinois 60607-7059\\
\par \vskip .3in

\end{center}

\begin{center}
L.A. Ferreira\footnotemark
{\footnotetext{Work supported in part by CNPq}}, J.F. Gomes$^{\,2}$
and A.H. Zimerman$^{\,2}$

\par \vskip .1in \noindent
Instituto de F\'{\i}sica Te\'{o}rica-UNESP\\
Rua Pamplona 145\\
01405-900 S\~{a}o Paulo, Brazil
\par \vskip .3in

\end{center}

\begin{center}
{\large {\bf ABSTRACT}}\\
\end{center}
\par \vskip .3in \noindent

We construct infinite sets of local conserved charges for
the conformal affine Toda model.
The technique involves the abelianization of the two-dimensional gauge
potentials satisfying the zero-curvature form of the equations
of motion.
We find two infinite sets of chiral charges and apart from two lowest spin
charges all the remaining ones do not possess chiral densities.
Charges of different chiralities Poisson commute among themselves.
We discuss the algebraic properties of these charges and use the fundamental
Poisson bracket relation to show that the charges conserved in time
are in involution.
Connections to other Toda models are established by taking particular limits.

\end{titlepage}

\sect{Introduction}
\label{sec:intro}
The completely integrable field theories are characterized by the existence
of infinite number of conserved charges in involution.
An important class of integrable theories is provided by the two-dimensional
Toda field theories in which integrability is signaled by the zero-curvature
form of the equations of motion.
It is of interest to reveal the complete integrability
of the Toda theories in its basic form constructing directly charges in
involution.
This has been done in the framework of the affine Toda (AT) model
based on the center-less Kac-Moody (KM) algebra.
There, the infinite set of local conserved
charges was constructed by the special technique of abelianizing the
two-dimensional gauge potentials associated to the zero curvature
equation \ct{OT85}.  In the case of $Sl(n)$ such charges where also
constructed, using other procedures, in \cite{Ruth}.

Within the context of Toda field theories one finds that the conformal affine
Toda (CAT) model \ct{BB,AFGZ} based on the full Kac-Moody algebra occupies a
special place due to its conformal invariance,
existence of ${\sf W}$-infinity symmetry \ct{deform,2boson}, the
soliton solutions \ct{CFGZ,soliton} and the fact that the AT as well
as the conformal Toda (CT)
model can be obtained from it by taking particular limits.
In view of the above one wishes to
investigate whether in the CAT model we have a strong form of complete
integrability.
This paper is devoted to  the explicit construction of the local, conserved
charges in involution for the CAT model.
The construction starts from equations of motion written in
the zero-curvature form and proceeds by gauge transforming potentials to
their abelian form.  We find, in this way, an intriguing structure of charges.
The two lowest charges have corresponding  chiral charge densities.
The remaining are chiral after integration but without having any chiral
charge density associated with them.
This explains why the local ${\sf W}$-symmetry structure of the CAT model
could be described by only two spin 1 and spin 2 chiral currents
\ct{hspin,deform,2boson}.
It is a surprising fact since the ${\sf W}$-algebra provided by these
two currents is the same for all CAT models irrespectively of the underlying
KM algebra.
In this paper the missing conservation laws associated with the structure of
KM algebra are uncovered.
However there are no extra local chiral densities extending the local ${\sf
W}$ symmetry.
This fact distinguishes the CAT model among the integrable conformal models.
For instance, the charges of the CT model are obtained from chiral densities
satisfying a ${\sf W}_N$ algebra intrinsically related to the Casimir
operators of the underlying simple Lie algebra \ct{ora}.

We prove using the fundamental Poisson bracket relations (FPR) that the
charges for CAT model conserved in time are in involution.
It is not clear yet whether the charges of given chirality are in involution
although it is quite trivial to see that charges of different chiralities
mutually Poisson commute.

The densities of the chiral charges have a special form, which allows one to
gauge fix the conformal symmetry obtaining the AT charges
constructed in \ct{OT85} (see also ref. \cite{Ruth}
for the SU(n) case) from the corresponding CAT charges.

In section \ref{sec:cat} we discuss the main properties of the CAT model.
We review some basic results and reformulate equations of motion in terms
of the special basis of KM algebra \ct{kostant,kac,OT85}.
This choice of basis proves to be essential in our construction.
The main results are given in section \ref{sec:charges} where by
successive gauge transformations we cast potentials into an
abelian form and find the corresponding conserved charges.
Their involution is discussed in section \ref{sec:invo} where use is
made of the fundamental Poisson bracket relations.  Finally in section
\ref{sec:examples}, we give explicit examples for $sl (2)$ and $sl(3)$
constructing the first few charges.  Section \ref{sec:disc} contains
discussion, and in appendix A we show how to construct a set of chiral charges
which are independent of the field $\nu$ of the CAT model.

\sect{The CAT model}
\label{sec:cat}
The Conformal affine Toda models (CAT) extend the Affine
Toda models (AT) by the introduction of two extra fields which make
the model conformally invariant \ct{BB,AFGZ}. For each simple Lie
algebra $\lie$ we associate a CAT model with the equations of motion
given by \ct{CFGZ}:
\br
\pa_{-} \pa_{+} \vp^a &=& \, \( q^{a} e^{ K_{ab}
\vp^b} - \,q^{0} l^{\psi}_{a}  e^{-  K_{\psi b} \vp^b } \) e^{ \eta}
\lab{todaone} \\ \pa_{-} \pa_{+} \eta &=& 0 \lab{todatwo} \\ \pa_{-}
\pa_{+} \nu &=&   {2 \o \psi^2} \, q^{0} e^{- K_{\psi b} \vp^{b}}
e^{ \eta} \lab{todathree}
\er
where $K_{ab}=2 \a_a.\a_b/{\a_b^2}$ is the Cartan Matrix of $\lie$,
$a,b=1,...$, ${\rm rank} \,\lie\equiv r$, $\psi$ is the highest root of
$\lie$, $K_{\psi b}=2 \psi .\a_b/{\a_b^2}$, $l^{\psi}_{a}$ are positive
integers appearing in the expansion ${\psi \over
\psi^{2}} = l^{\psi}_{a} {\a_{a} \over \a^{2}_{a}}$, where $\a_a$
are the simple roots of $\lie$ and $q^a$ and $q^0$  are coupling
constants.

These equations are invariant under conformal transformations:
\br
x_{+} \rightarrow \tilde{x}_{+} = f(x_{+})
 \, \, \, , \, \, \, \,
x_{-} \rightarrow \tilde{x}_{-} = g(x_{-}) \lab{ge}
\er
and
\br
e^{-\vp^a (x_+,x_-)} &\to &
e^{-\tilde{\vp}^a(\tilde{x}_+,\tilde{x}_-)} = e^{-\vp^a (x_+,x_-)}
\lab{fi} \\
e^{-\nu (x_+,x_-)} &\to &
e^{-\tilde{\nu}(\tilde{x}_+,\tilde{x}_-)}= ({df \over dx_+})^{B}
({dg \over dx_-})^{B} e^{-\nu (x_+,x_-)} \lab{ni}\\
e^{-\eta (x_+,x_-)} &\to &
e^{-\tilde{\eta}(\tilde{x}_+,\tilde{x}_-)} =({df \over
dx_+}) ({dg \over dx_-}) e^{-\eta
(x_+,x_-)}   \lab{mi} \er
where $f$ and $g$ are analytic functions and $B$ is an arbitrary
constant. Therefore $e^{\varphi^a}$ are scalars under conformal
transformations and $e^{-\nu} $ and $e^{-\eta}$ are primary
fields of conformal weights $(B,B)$ and $(1,1)$
respectively \ct{CFGZ}.

The equations of motion \rf{todaone}-\rf{todathree} can be written as
a zero curvature condition:
\be
\pa_{+} A_{-} - \pa_{-} A_{+} + \lb A_{+} , A_{-} \rb = 0
\lab{zc}
\ee
where the potentials $A_{\pm}$ lie on the affine Kac-Moody algebra $\alie$
associated to $\lie$
\br
A_{+} = {1\over 2}\pa_{+} \Phi + e^{ad \Phi /2} {\cal E}_{+} \, \, \,
, \, \, \, A_{-} = - {1\over 2}\pa_{-} \Phi + e^{-ad \Phi /2} {\cal
E}_{-} \lab{potentials}
\er
and
\br
\Phi &=& \sum_{a=1}^{r} \varphi^a H_a^0 + \eta {\hat \rho} +
\nu C  \lab{Phi} \\
{\cal E}_{+} &=& \sum_{a=1}^{r}  \sqrt{l^{\psi}_a} E_{\a_a}^0 +
 E_{-\psi}^1 \, \, \, , \, \, \, {\cal E}_{-} = \sum_{a=1}^{r} {q^a
\o \sqrt{l^{\psi}_a}}  E_{-\a_a}^0 + q^0 E_{\psi}^{-1}
\lab{gauge}
\er
where
\be
\rh = 2 {\hat \delta}. H^0 + h D
\lab{rho}
\ee
with ${\hat \delta}={1\over 2}\sum_{\a > 0}{\a \over {\a^2}}$, is the
generator used to perform the so called principal gradation of an
affine Kac-Moody algebra.

The generators $H_a^n$, $E_{\a}^n$, $C$ and $D$ constitute the Chevalley
basis of $\alie$ and their commutation relations are defined in \rf{km}.
There exists however another basis for $\alie$ which will prove very useful
in the construction  of the conserved charges of the CAT model
\ct{kostant,OTU93}. In that basis the element ${\cal E}_{+}$ appearing in
the gauge potentials \rf{potentials} plays a crucial role.
One can grade $\alie$ using the operator \rf{rho} as
\be
\alie = \oplus \; \alie_m
\ee
where ${m \in \IZ}$ and such that
\be
\lb \rh \, , \, \alie_m \rb = m \; \alie_m
\ee
The subalgebra $\hlie$ of $\alie$ commuting,
modulo the central term $C$, with ${\cal E}_{+}$ is a Heisenberg
subalgebra with generators $E_M \in \alie_M$, where $M$ are the
exponents of $\alie$, satisfying
\be
\lb E_M \, , \, E_N \rb = {C \o h} Tr(E_M E_N) M \d_{M+N,0}
\ee
These exponents have a period equal
to the Coxeter number $h$ of $\lie$ and for $\alie$ being an affine
Kac-Moody algebra (not twisted) they have the form $M \equiv m_a + n
h$ where $n$ is an integer and $m_a$, $a=1,2, \ldots, r$ are the
exponents of the simple Lie algebra $\lie$. In addition these
exponents possess the symmetry $M \rightarrow -M$.
In particular, unity is always an exponent and it
follows that
\be
E_{1} = {\cal E}_{+} \, \, \, , \, \, \,  E_{-1} =
\sum_{a=1}^{r} \sqrt{l^{\psi}_a} E_{-\a_a}^0 + E_{\psi}^{-1}
\lab{e1}
\ee
The complement $\flie$ of $\hlie$ in $\alie$ is such that the
dimension of $\flie_m \subset \alie_m$ for $m \ne 0$ is equal to the
rank of $\lie$ ($\equiv r$). The subspace $\flie_0$ has dimension
$r+2$ and is generated by $C$, $D$ and $H_a^0$, $a=1,2, \ldots, r$.
Except for the extra
two generators of $\flie_0$, the basis, $F^a_m$, of
$\flie_m$  can be chosen such that
\be
\lb E_M \, , \, F^a_n \rb =
\gamma^a \cdot v_{[M]} F^a_{M+n} \lab{ef}
\ee
where $\gamma^a$ and $v_{[M]}$ are vectors in the root space of
$\lie$ and $[M]$ means $M$ modulo $h$ \ct{kac,OTU93}.

\sect{The Construction of the Charges}
\label{sec:charges}
We now show how to construct the conserved charges for the
CAT models.
In the case of the AT models this was done by rotating the gauge
potentials into an abelian subalgebra such that the
commutator term in \rf{zc} vanishes \ct{OT85}. One then obtains, by
imposing suitable boundary conditions on one of the space time
variables, let us say $x$,  that the quantity $\int dx A_x$ is
constant in the other variable, let us say $t$. Such procedure
proved to be very powerful in the case of the AT models \ct{OT85}
where the underlying algebra is a loop algebra.  In the case of the
CAT models the situation is a bit more complicated
due to the central term of the Kac-Moody algebra.
However,  such fact does not prevent us from constructing
conserved charges.
In fact, it makes things even more interesting.
We can rotate the right/left components of the gauge potentials into
the negative/positive abelian subalgebras of the Heisenberg
algebra $\hlie$.
The structure of those potentials is such that the zero curvature form of the
equations of motion leads to the conservation of the densities
of the chiral components of the energy-momentum tensor, and also
of the spin one current related to the free field $\eta$.
This contrasts with the AT model, where only the integrated E-M tensor is
conserved \ct{OT85}. This sheds some light on the
connection between the conformal invariance of the model and the
central extension of the KM algebra. Another interesting
aspect is that the higher chiral conserved charges are integrals of
non chiral densities. Therefore the conservation laws of the CAT
model are not really described by a ${\sf W}$ algebra like in the case of
the CT models where one has chiral densities generating a ${\sf W}_N$
algebra. However, as we have shown in \ct{hspin,deform,2boson} part of
the conservation laws for the CAT model leads to a $\Winf$
algebra.

We now show that there are two infinite sets of chiral conserved charges.
Each right (left) chiral charge is associated with a generator $E_M$ for
$M<0$ ($M>0$). In fact, the charges will be conserved
in time $t$, $x_{+}$ or $x_{-}$ depending upon the boundary conditions one
imposes on the fields.
In this section we will work out the conservation laws using light
cone variables.
In section \ref{sec:invo} we show that the charges conserved in time
are in involution.

Let us start with the right charges. We first
perform a gauge transformation
\be
A_{\pm}
\rightarrow g^{-1} A_{\pm} g + g^{-1} \pa_{\pm} g \lab{gt}
\ee
with $g = e^{\Phi /2}$. We then get
\br
A_{+} &\rightarrow & A_{+}^R = \pa_{+} \Phi +  E_1 \nonu \\
A_{-} &\rightarrow & A_{-}^R = e^{-\Phi} {\cal E}_{-} e^{\Phi}
\lab{rp}
\er
The idea now is to perform a second gauge transformation to rotate
the potential $A_{+}^R$ as
\br
A_{+}^R \rightarrow \ca_{+}^R &=&  g_R^{-1} A_{+}^R g_R +
g_R^{-1} \pa_{+} g_R \nonu \\
&=& E_1 + \pa_{+} \eta \, \rh + \sum_{M>0} \ca_{+}^{R,(M)} E_{-M}
\lab{gtaplus}
\er
where $g_R$ is an exponentiation of the generators of $\alie$ with
negative eigenvalues of $\rh$ given in \rf{rho} \ct{paunov}. In fact, since
$E_{-M}$ for $M>1$ commutes with $E_1$ we can take $g_R$ to be an
exponentiation of $E_{-1}$ and the generators of $\flie_{-m}$ for
$m>0$. To show that \rf {gtaplus} is possible, let us write
\br
\ca_{+}^R &=&
\sum_{m=1}^{-\infty} \ca_{+}^{R,(m)}
\, \, \, , \, \, \,  \mbox{\rm with} \, \, \ca_{+}^{R,(m)} \in
\alie_{m} \\
g_R &=&
e^{(\chi E_{-1} + \sum_{m>0} \zeta_{-m})} \, \, \, , \, \, \,
\mbox{\rm where} \, \, \zeta_{-m} \in \flie_{-m}
\er
Then
\br
\ca_{+}^{R,(1)} &=& E_1 \nonu \\
\ca_{+}^{R,(0)} &=& \lb E_1 \, , \, \chi \, E_{-1} + \zeta_{-1} \rb +
\pa_{+} \Phi \nonu \\
\ca_{+}^{R,(-1)} &=& \lb E_1 \, , \,  \zeta_{-2}
\rb + \lb \pa_{+} \Phi \, , \, \chi \, E_{-1} + \zeta_{-1} \rb +
{1 \o 2!} \lb \chi \, E_{-1} + \zeta_{-1} \, , \, \lb \chi \, E_{-1}
+  \zeta_{-1} \, , \, E_1 \rb \rb \nonu\\
& & + \pa_{+} \chi \, E_{-1} + \pa_{+} \zeta_{-1} \nonu \\
\vdots & & \vdots \nonu \\
\ca_{+}^{R,(-m)} &=& \lb E_1 \, , \,
\zeta_{-m-1} \rb + X_{-m}
\lab{expansion}
\er
where $X_{-m}$ depends on $\zeta_{-n}$ for $n\leq m$ only.

One can observe that the $r$ parameters in $\zeta_{-m-1}$ can be
chosen in such a way as to cancel the $r$ components of $\ca_{+}^{R,(-m)}$
lying in $\flie_{-m}$. The exception is on the level zero where we
have also to fix $\chi$, in addition to $\zeta_{-1}$, in order to
remove the $r+1$ components of $\ca_{+}^{R,(0)}$ lying in the direction of
$H_a^0$ ($a=1,2,...r$) and $C$. Then $\ca_{+}^{R,(0)}$ will
be just $\pa_{+} \eta \, \rh$. Therefore it is possible to gauge
transform $A_{+}^R$ into the form \rf{gtaplus}.

For the potential $A_{-}^R$ we get
\br
A_{-}^R \rightarrow \ca_{-}^R &=&  g_R^{-1} A_{-}^R g_R + g_R^{-1}
\pa_{-} g_R  \nonu \\
&=& \sum_{M>0} \ca_{-}^{R,(M)} E_{-M} + \sum_{a=1}^r \sum_{m>0}
b_{-}^{R,(a,m)} F_{-m}^a
\lab{gtaminus}
\er
The components of $\ca_{-}^R$ in the direction of $F_{-m}^a$, as we
show below, vanish as a consequence of the equations of motion. The
zero curvature condition \rf{zc} for the gauge transformed
potentials can be split into three parts, one in the direction
of $\rh$, other lying on the Heisenberg subalgebra and another on
its complement. So one gets
\br
0 &=&  \pa_{+} \pa_{-} \eta \, \rh \lab{zc0} \\
0 &=&  \sum_{M>0} \( \pa_{+} \ca_{-}^{R,(M)} -
\pa_{-} \ca_{+}^{R,(M)} \) E_{-M}  \nonu \\
 &+& \llbrack E_1 + \pa_{+} \eta
\, \rh + \sum_{M>0} \ca_{+}^{R,(M)} E_{-M} \, , \,  \sum_{M>0}
\ca_{-}^{R,(M)} E_{-M} \rrbrack   \lab{zc1}\\
0 &=&  \sum_{m>0} \pa_{+}
b_{-}^{R,(a,m)} F_{-m}^a +  \llbrack E_1 + \pa_{+} \eta \, \rh +
\sum_{M>0} \ca_{+}^{R,(M)} E_{-M} \, , \,  \sum_{a=1}^r \sum_{m>0}
b_{-}^{R,(a,m)} F_{-m}^a \rrbrack \lab{zc2}
\er
In eq. \rf{zc2} the term with highest eigenvalue of $\rh$ is
\be
\lb E_1 \, , \, \sum_{a=1}^r b_{-}^{R,(a,1)} F_{-1}^a \rb =0
\lab{b1}
\ee
Since there are no generators of $\flie$ commuting with $E_1$ (see
\rf{ef}) one concludes that $b_{-}^{R,(a,1)}=0$. Now, the term with
highest eigenvalue of $\rh$ in \rf{zc2} becomes $\lb E_1 \, , \,
\sum_{a=1}^r b_{-}^{R,(a,2)} F_{-2}^a \rb =0$ and for the same
reason $b_{-}^{R,(a,2)}=0$. So, following this reasoning
one concludes that the second term on the right hand side of
\rf{gtaminus} vanishes.
Therefore we have rotated the two components of the gauge potential
into the Heisenberg subalgebra (except for the term $\pa_{+} \eta \,
\rh$ in $\ca_{+}^R$). One observes that the only non vanishing term,
involving $E_1$, in the commutator in \rf{zc1} is given by
\be
\lb E_1 \, , \, \ca_{-}^{R,(1)} E_{-1} \rb \sim C
\lab{am1}
\ee
Since this is the
only term in the direction of $C$ in \rf{zc1} one concludes that
$\ca_{-}^{R,(1)} = 0$.
Hence we have proved that
\be
\ca_{-}^R = \sum_{M>1} \ca_{-}^{R,(M)} E_{-M}
\ee
In consequence the terms proportional to $E_{-1}$
in \rf{zc1} lead to
\be
\pa_{-} \ca_{+}^{R,(1)} = 0
\lab{aplusr1}
\ee
$\ca_{+}^{R,(1)}$ corresponds, in fact,  to a chiral component of the
CAT model stress tensor. Notice that the central term played a
crucial role in making such local density chiral. In addition
\rf{zc0} leads to
\be
\pa_{-} J^R = 0  \, \, \, , \, \, \,
\mbox{\rm with} \, \, J^R \equiv \pa_{+} \eta
\lab{jr}
\ee
The chiral densities $J^R $ and $\ca_{+}^{R,(1)}$ are respectively the
spin $1$ and $2$ fields shown, in \ct{hspin}, to be the only
remaining Kac-Moody currents in the reduction of the two-loop WZNW
model \ct{AFGZ,FGSZ}, and used to construct an $\Winf$
for the CAT model \ct{deform,2boson}.

After those considerations what is left of \rf{zc1} is
\be
(\pa_{+} - M J^R ) \ca_{-}^{R,(M)} - \pa_{-} \ca_{+}^{R,(M)} =0 \, \,
\, , \, \,  \, \mbox{\rm for $M>1$}
\lab{conservationr1}
\ee
These equations are in fact the conservation laws for the CAT model.
They can be rewritten as
\be
\pa_{+} a_{-}^{R,(M)}= \pa_{-} a_{+}^{R,(M)}\, \,
\, , \, \,  \, \mbox{\rm for $M>1$}
\lab{conservationr2}
\ee
where we have introduced
\br
a_{+}^{R,(M)} &=& e^{-M \int^{x_{+}} dy_{+} J^R} \ca_{+}^{R,(M)}\nonu\\
a_{-}^{R,(M)} &=& e^{-M \int^{x_{+}} dy_{+} J^R} \ca_{-}^{R,(M)}\, \,
\, , \, \,  \, \mbox{\rm for $M>1$}
\lab{abpot}
\er
Therefore by imposing suitable boundary conditions on the fields it
follows from \rf{jr} and \rf{conservationr2} that the charges
\be
Q_M^R = \int dx_{+} e^{-M \int^{x_{+}} dy_{+} J^R}  \ca_{+}^{R,(M)}
\lab{chargesR}
\ee
are chiral
\be
\pa_{-} Q_M^R = 0
\ee
Notice that $e^{-M \int^{x_{+}} dy_{+} J^R}$ is
$x_{-}$ independent, but the quantities $\ca_{+}^{R,(M)}$ are not.
Therefore the densities of the chiral charges \rf{chargesR}
are not chiral themselves. Hence we have here a situation
different from that of conformal field theories with extended
symmetry. There, the symmetries are described by in general a
${\sf W}$-algebra and the densities are chiral (like in the CT models
for instance).
It would be interesting to study the algebra of these non chiral
densities.

The gauge potentials $\ca_{\pm}^R$ can in fact be rotated into the
abelian subalgebra of $\hlie$ generated by $E_{-M}$  for $M<0$.
This is done by two gauge transformations. Denoting
\be
g_1(x_{+}) = e^{-\rh \int^{x_{+}} J^R dy_{+}} \quad {\rm  and}
\quad g_2 (x_{+}) = \exp ( - E_1 \int^{x_{+}} e^{ \int^{y_{+}}
J^R dz_{+}}dy_{+})
\ee
we get
\br
\ca_{+}^R \rightarrow  {\bar \ca}_{+}^R &=&  g_1^{-1} \ca_{+}^R g_1 +
g_1^{-1} \pa_{+} g_1 \nonu \\
&=& e^{\int^{x_{+}} J^R dy_{+}} E_1 +
\sum_{M>0} a_{+}^{R,(M)} E_{-M}
\er
where now we extend the definition \rf{abpot} of $a_{+}^{R,(M)}$ for $M=1$. In
addition
\br
{\bar \ca}_{+}^R \rightarrow  a_{+}^R &=&  g_2^{-1} {\bar \ca}_{+}^R g_2 +
g_2^{-1} \pa_{+} g_2 \nonu \\
&=& a_{+}^{R(0)} C + \sum_{M>0} a_{+}^{R,(M)} E_{-M}
\er
where
\be
a_{+}^{R(0)} = {1\o h}Tr(E_1 E_{-1}) \ca_{+}^{R,(1)} e^{-\int^{x_{+}}
J^R dz_{+}} \int^{x_{+}} e^{\int^{y_{+}} J^R dw_{+}} dy_{+}
\ee
The other component of the gauge potential transforms as
\br
\ca_{-}^R \rightarrow  {\bar \ca}_{-}^R &=&  g_1^{-1} \ca_{-}^R g_1 +
g_1^{-1} \pa_{-} g_1 \nonu \\
&=&  \sum_{M>1} a_{-}^{R,(M)} E_{-M}
\er
where $a_{-}^{R,(M)}$ was defined in \rf{abpot}. Finally
\br
{\bar \ca}_{-}^R \rightarrow  a_{-}^R &=&  g_2^{-1} {\bar \ca}_{-}^R  g_2 +
g_2^{-1} \pa_{+} g_2 \nonu \\
&=& {\bar \ca}_{-}^R
\er
since $E_1$ commutes with all $E_{-M}$ for $M<0$ and $J^R$ is $x_{-}$
independent.

Therefore the gauge potentials $a_{\pm}^R$ are abelian and satisfy the
zero curvature condition:
\be
\pa_{+} a_{-}^R - \pa_{-} a_{+}^R = 0
\ee
Notice that the only non-local quantity in these potentials is
$a_{+}^{R,(0)}$, but it is harmless due to the chiral character of the
quantities involved.
Since $\int^{x_{+}} J^R dy_{+} = \eta_{+} (x_{+})$, where
$\eta (x_{+}, x_{-})= \eta_{+} (x_{+})+ \eta_{-} (x_{-})$ is a solution
of the equations of motion, the exponential terms appearing in
$a_{\pm}^R$ are local quantities and depend only on the free field $\eta$.
That is an interesting fact and it is a consequence of the conformal symmetry.
As we have shown in \ct{CFGZ} the dynamics of the CAT model on a space-time
$(x_{+},x_{-})$ is the same as that of the AT model (except for the
passive field $\nu$) on a space time $({\tilde x}_{+},{\tilde
x}_{-})$ where $x$'s and ${\tilde x}$'s are related by the
transformation
\be
{\tilde x}_{+} = \int^{x_{+}} dy_{+}
e^{\eta_{+}(y_{+})} \, \, \, , \, \, \,  {\tilde x}_{-} =
\int^{x_{-}} dy_{-} e^{\eta_{-}(y_{-})}
\lab{just it}
\ee
So, for every regular solution of the $\eta$ field the CAT model is an AT
model in a different space-time.

One can easily convince oneself, by studying \rf{expansion}, that the
quantity $\ca_{+}^{R,(M)}$ is a polynomial in the derivatives of the
CAT model fields and each term in it contains exactly $M+1$
derivatives w.r.t.  $x_{+}$ and no derivatives w.r.t. $x_{-}$.
 Therefore, under the  transformation
\rf{just it} we have from \rf{chargesR}
\be
\int \( e^{\int^{x_{+}} dy_{+}
J^R}dx_{+}\) e^{-(M+1) \int^{x_{+}} dy_{+} J^R}
\ca_{+}^{R,(M)}(x_{+}, x_{-}) \rightarrow  \int d{\tilde x}_{+}
{\tilde \ca}_{+}^{R,(M)}({\tilde x}_{+}, {\tilde x}_{-})
\ee
It is a remarkable fact that the term $e^{-M \int^{x_{+}} dy_{+}
J^R}$  comes exactly with the right form to make such a
transformation to the AT charges possible. The density ${\tilde
a}_{+}^{R,(M)}({\tilde x}_{+}, {\tilde x}_{-})$ is now
$\eta$-independent, although it may be $\nu$-dependent. In the appendix A we
show how to construct a set of densities which are $\nu$ independent. Those
relate to the densities for the AT model which were  constructed in \ct{OT85},
and also in \cite{Ruth}.

We now discuss the construction of the left charges for the CAT
models. We first perform a global gauge transformation on the gauge
potentials \rf{potentials} with the group element  $\Omega =
\exp \(\sum_{b=1}^r\omega_b 2 \lambda_b \cdot H^0/\a^2_b + \omega_0
D \)$, where $\lambda_a$ are the fundamental weights of $\lie$,
$\omega_a = \log (q_a/l_a^{\psi})$ and $\omega_0 = \sum_{b=1}^r
n_b^{\psi}\omega_b + \log q_0$ with $n_b^{\psi}$ being the integers
in the expansion $\psi = n_b^{\psi} \a_b$. We then get
\br
A_{+} \rightarrow {\bar A}_{+} &=& \Omega A_{+} \Omega^{-1} =
{1\over 2}\pa_{+}\Phi + e^{ad \Phi /2} {\cal E}_{+}^{L}\\
A_{-} \rightarrow  {\bar A}_{-} &=& \Omega A_{-} \Omega^{-1} =
 - {1\over 2}\pa_{-} \Phi + e^{-ad \Phi /2} E_{-1}
\lab{global}
\er
where
\be
{\cal E}_{+}^{L} = \sum_{a=1}^{r}  {q_a \o \sqrt{l_a^{\psi}}}
E_{\a_a}^0 + q_0 E_{-\psi}^1
\ee
Next, we perform the gauge transformation \rf{gt} with $g=e^{-\Phi
/2}$. We get
\br
{\bar A}_{+} &\rightarrow & A_{+}^L = e^{\Phi} {\cal E}_{+}^L
e^{-\Phi} \nonu \\
{\bar A}_{-} &\rightarrow & A_{-}^L = -\pa_{-} \Phi +  E_{-1}
\lab{lp}
\er
We now rotate these potentials into the Heisenberg subalgebra with
a third gauge transformation
\br
A_{\pm}^L \rightarrow \ca_{\pm}^L =  g_L^{-1} A_{\pm}^L g_L +
g_L^{-1} \pa_{\pm} g_l
\er
with $g_L = e^{(\xi E_{1} + \sum_{m>0} \zeta_{m})}$ and $\zeta_{m}
\in  \flie_{m}$. Following the same procedure we used in the case
of the right charges we obtain that
\br
\ca_{-}^L &=& E_{-1} - \pa_{-} \eta \rh + \sum_{M>0} \ca_{-}^{L,(M)}
E_{M} \nonu\\
\ca_{+}^L &=& \sum_{M>1} \ca_{+}^{L,(M)} E_{M}
\er
Then by performing two consecutive gauge transformations with ${\bar g}_1
= \exp (\rh \int^{x_{-}} J^L dy_{-} )$ and then ${\bar g}_2
= \exp (-E_{-1} \int^{x_{-}} \exp ( \int^{y_{-}} J^L dz_{-})dy_{-})$,
with $J^L$ defined in \rf{jl}, we get
\br
\ca_{-}^L \rightarrow a_{-}^L &=&  a_{-}^{L,(0)} C + \sum_{M>0} a_{-}^{L,(M)}
E_{M} \nonu\\
\ca_{+}^L \rightarrow a_{+}^L &=&  \sum_{M>1} a_{+}^{L,(M)} E_{M}
\er
with
\br
a_{-}^{L,(0)} &=& -{1\o h} Tr(E_1 E_{-1}) \ca_{-}^{L,(1)}
e^{-\int^{x_{-}} J^L dz_{-}} \int^{x_{-}} e^{\int^{y_{-}}
J^L dw_{-}} dy_{-}\nonu\\
a_{\pm}^L &=& \exp (-M \int^{x_{-}} J^L dy_{-}) \ca_{\pm}^L
\er
So, the gauge potentials are abelian and satisfy
\be
\pa_{+} a_{-}^L - \pa_{-} a_{+}^L = 0
\ee
and lead to the conservation laws
\br
\pa_{+} a_{-}^{L,(1)} &=& 0 \\
\pa_{+} J^L &=& 0  \, \, \, , \, \, \,
\mbox{\rm with} \, \, J^L \equiv \pa_{-} \eta
\lab{jl}\\
\pa_{+} Q_M^L &=& 0 \, \, \, , \, \,  \, \mbox{\rm for $M>1$}
\er
where
\be
Q_M^L = \int dx_{-}  a_{-}^{L,(M)}
\lab{chargesL}
\ee
We recognize in the above the similar structure as the one we previously
constructed for the right charges.

Notice that the densities $a_{+}^{R,(M)}$  ($a_{-}^{L,(M)}$) are functions  of
$x_{+}$ ($x_{-}$) derivatives of the fields only. For Lagrangeans which are
quadractic in time derivatives of the fields, it follows that the Poisson
bracket between $x_{+}$ and $x_{-}$ derivatives of the fields vanishes.
Therefore the right and left charges Poisson commute. So, the algebra of the
chiral charges split into two commuting isomorphic subalgebras.

\sect{Involution of Charges}
\label{sec:invo}
In this section we will use the fundamental Poisson bracket
relation (FPR) and abelianization to verify that the $x$-component charges
obtained above are indeed in involution.
We will follow closely approach of references \ct{Fad-LH}
and \ct{OT85}.
Let us recall some main steps of this formalism.
The basic role is played by the FPR relation constructed in \ct{AFGZ} for the
CAT model:
\be
\{ A_x (x) \stackrel{\otimes}{,}  A_x (y) \}_{PB} = - \h \delta (x-y)\;
\biggl\lbrack \, \IP \,, \, 1 \otimes A_x (x) + A_x (x) \otimes 1 \,\biggr
\rbrack
\lab{fpr}
\ee
where $A_x = A_{+} + A_{-}$, with $A_{\pm}$ given in \rf{potentials} and
\be
\IP \,=\, \IC_{+} - \IC_{-}
\lab{pandc}
\ee
and $\IC_{\pm}$ are given by
\br
\IC_{+} &=& \sum_{m = 1}^{\infty} \sum^{r}_{a=1}
\eta^{ab} H^m_a \otimes H^{-m}_b + \sum_{\a >0} { \a^2 \o 2} E^0_{\a} \otimes
E^0_{-\a} \lab{casplus}\\
&+ & \sum_{m = 1}^{\infty} \sum_{\a >0} { \a^2 \o 2} \(
E^m_\a \otimes E^{-m}_{- \a} + E^m_{- \a} \otimes E^{-m}_{\a}\)
\nonumber
\er
and $\IC_{-}$ is obtained from $\IC_{+}$ by the interchange of left and right
entries, and $\eta^{ab}$ is the inverse of $\eta_{ab}= {2\o \a_a^2} K_{ab}$.

Due to the commutator term in \rf{zc}, the quantities $\int_{-l}^l A_x dx$ are
not conserved in time. However denoting $a_x^{R/L} = a_{+}^{R/L} +
a_{-}^{R/L}= \sum_{M>0} a_x^{R/L,(M)}E_{\pm M}$, with $a_{\pm}^{R/L}$
constructed in section \ref{sec:charges},
one obtains that the quantities
\be
P^{R/L}_M \equiv \int_{-l}^l a_x^{R/L,(M)} \; dx
\lab{xcharges}
\ee
are conserved in time when periodic boundary conditions are imposed on the
fields at $x=\pm l$ . However the bracket involving $a_x^{R/L}$ is not easy
to evaluate directly. Since they are related to $A_x$ via gauge transformations
as shown in section \ref{sec:charges}, we will make use of the gauge invariant
quantities ${\rm Tr} U^n$ with $U$ being the path ordered exponential
\be
U = T \exp \lb - \int_{-l}^{l} A_x (x) dx \rb  \lab{umatrix}
\ee
Indeed, using the zero curvature condition \rf{zc} and the non abelian
version of Stoke's theorem \ct{OT85} one gets that ${\rm Tr} U^n$ are
conserved in time.

{}From FPR \rf{fpr} we can prove that
\be
\{ A_x (x) \stackrel{\otimes}{,}  U \}_{PB} = - \lb A_x (x) \otimes 1\, , \,
Q  (x) \rb \, - \, \partder { Q (x)}{x}
\lab{aq}
\ee
where
\be
Q (x) = 1 \otimes Te^{-\int_{-l}^x A_x (y) dy} \, \h \IP \,
1 \otimes Te^{-\int^{l}_x A_x (y) dy}
\lab{qdef}
\ee
Relation \rf{aq} can be used to prove another useful identity
\be
\{ A_x (x) \, , \,   {\rm Tr} U^m  \}_{PB} = - \lb \pa_x + A_x (x) \, , \,
{\hat Q}_m  (x) \rb
\lab{aqhat}
\ee
with
\be
{\hat Q}_m  (x) = m {\rm Tr}_r \( Q (x) 1 \otimes U^{m-1} \)
\lab{qhatdef}
\ee
One can now show that the Poisson bracket
$ \{ T \exp \( - \int_{-l}^{l} A_x (y)dy \)  \, , \,   \Tr U^m  \}_{PB}$
is equal to an integration of a total derivative and therefore
vanishes when we assume the appropriate periodic boundary conditions for the
interval $ \lb -l , l \rb$.
Hence by this standard argument it follows that the quantities
$\Tr U^m $ are in involution for all integers $m$
\be
\{ {\rm Tr} U^n \, , \,   {\rm Tr} U^m  \}_{PB} = 0
\lab{unum}
\ee
Since ${\rm Tr} U^n$ are gauge invariant, they are equal to ${\rm Tr}
(u^{R/L})^n$, where $u^{R/L}= T \exp \lb - \int_{-l}^{l} a^{R/L}_x (x) dx
\rb$. But as we have shown in section \ref{sec:charges} the potentials
$a_x^{R/L}$ are abelian and so path ordering is not essential. Therefore
the conserved quantities ${\rm Tr} U^n$ are local. In fact, they are
functionally related
to the charges \rf{xcharges}.
{}From this we conclude that the integrals of $x$-components of
$a_{\pm}^{R,(M)}$, i.e. the charges \rf{xcharges}, must also be in involution
as a consequence of \rf{unum}.

\sect{Examples}
\label{sec:examples}

The commutation relations for an affine Kac-Moody algebra in the
Chevalley basis are given by
\br
\lb H_a^m \, , \, H_b^n \rb &=& C {2\o \a_a^2} K_{ab} m \d_{m+n,0} \nonu\\
\lb H_a^m \, , \, E_{\pm \a}^n \rb &=& \pm K_{\a a} E_{\pm \a}^{m+n} \nonu \\
\lb E_{\a}^m \, , \, E_{-\a}^n \rb &=&  l_a^{\a} H_a^{m+n} +
C {2\o \a^2} m \d_{m+n,0} \nonu\\
\lb E_{\a}^m \, , \, E_{\b}^n \rb &=& \epsilon (\a , \b )
E_{\a + \b}^{m+n} \, \, \, \mbox{\rm if $\a + \b$ is a root of $\lie$}
\nonu\\
\lb D \, , \, H_b^m \rb &=& m H_b^m \nonu\\
\lb D \, , \, E_{\a}^m \rb &=& m  E_{\a}^m
\lab{km}
\er
where $K_{\a a}= 2 \a . \a_a / \a_a^2 = n_b^{\a} K_{ba}$, with $n_a^{\a}$
and $l^{\a}_a$ being the integers in the expansions
$\a = n^{\a}_a \a_a$ and $\a /\a^2 = l^{\a}_a \a_a / \a_a^2$, and
$\epsilon (\a , \b )$ being structure  constants.
We now give explicit expressions for the first two non trivial charges for
the case of the $Sl(2)$ and $SL(3)$ CAT models.

\subsection{The Example of ${\bf Sl(2)}$}
\label{sec:sl2}

The simple algebra $Sl(2)$ has just one positive (simple) root and
its Cartan matrix is $K=2$. The equations of motion are those
given in \rf{todaone}-\rf{todathree} with $K_{ab}=K_{\psi b}=2$,
$l_a^{\psi}=1$ and we shall normalize $\psi^2=2$. The basis discussed
in section \ref{sec:cat} in this case is given by
\br
E_{2m+1} &=& E_{\a}^m + E_{-\a}^{m+1} \\
F_{2m+1} &=& E_{\a}^m -E_{-\a}^{m+1} \\
F_{2m} &=& H^{m} - {1\o 2} C \d_{m,0}
\er
in addition to the generators $C$ (central term) and $\rh = {1\o 2}
H^0 + 2 D$. The commutation relations are
\br
\lb E_{2m+1} \, , \, E_{2n+1} \rb &=& C (2m+1) \d_{m+n+1,0} \\
\lb E_{2m+1} \, , \, F_{2n+1} \rb &=& - 2 F_{2(m+n+1)} \\
\lb E_{2m+1} \, , \, F_{2n} \rb &=& - 2 F_{2(m+n)+1} \\
\lb F_{2m+1} \, , \, F_{2n+1} \rb &=& - C (2m+1) \d_{m+n+1,0} \\
\lb F_{2m+1} \, , \, F_{2n} \rb &=& - 2 E_{2(m+n)+1} \\
\lb F_{2m} \, , \, F_{2n} \rb &=& C 2 m \d_{m+n,0}
\er
and they are all eigenvectors of $\rh$
\br
\lb \rh \, , \, E_{2m+1} \rb &=& (2m+1) E_{2m+1} \\
\lb \rh \, , \, F_{2m+1} \rb &=& (2m+1) F_{2m+1} \\
\lb \rh \, , \, F_{2m} \rb &=& 2m F_{2m}
\er
Using the procedure of section \ref{sec:charges} one can calculate
the charges. The first two right charge densities are given by
\br
\ca_{+}^{R,(1)} &=&   {1\o 2}(\pa_{+} \vp )^2 -
{1\o 2} \pa_{+}^2 \vp
+ {1\o 2} \pa_{+} \vp \pa_{+} \eta +  \pa_{+} \eta \pa_{+} \nu -
\pa_{+}^2 \nu \\
\ca_{-}^{R,(1)} &=& 0
\er
and
\br
a_{+}^{R,(3)} &=& {1\o 8} e^{-3\int^{x_{+}}J^R}\( -(\pa_{+} \vp )^4  +
\pa_{+} \vp
\pa_{+}^3 \vp + {1\o 2} \pa_{+}^2 \eta (\pa_{+} \vp )^2 - {5\o 2}
(\pa_{+} \eta \pa_{+} \vp )^2 \right. \nonu \\
 & & \left. + (\pa_{+} - 3 \pa_{+} \eta ) \( - {1\o 3} (\pa_{+} \vp )^3 -
A{2\o 3} \pa_{+} \nu (\pa_{+} \vp )^2 - {3\o 2} \pa_{+} \eta
(\pa_{+} \vp )^2 \) \) \\
a_{-}^{R,(3)} &=& {1\o 8} e^{-3\int^{x_{+}} J^R}\( q_0 e^{-2 \vp + \eta} \(
{1\o 3} (\pa_{+} \vp )^2 + \pa_{+} \eta \pa_{+} \vp + {4\o 3}
\pa_{+} \nu \pa_{+} \vp - \pa_{+}^2 \vp \) \right. \nonu\\
 & & \left. + q_1 e^{2 \vp +\eta} \( - (\pa_{+} \vp )^2 - \pa_{+} \eta
\pa_{+} \vp - {4\o 3} \pa_{+} \nu \pa_{+} \vp + \pa_{+}^2 \vp \) \)
\er
As explicit calculation has shown the integral of charge densities
$a_{+}^{R,(k)}\; k=1,3,5$ becomes $\nu$-independent under appropriate boundary
conditions. This is related to the fact that it is possible to construct
$\nu$-independent densities, as shown in the appendix A.
Also we should point out that the charges obtained from $a_{+}^{R,(k)}\; k=
3,5,... $ correspond to taking the AT charges and replacing the space time
coordinates by \rf{just it}

\subsection{The Example of ${\bf Sl(3)}$}
\label{sec:sl3}

The Cartan matrix for $SL(3)$ is given by $K_{11}=K_{22}=2$
and $K_{12}=K_{21}=-1$ and $l^{\psi}_{a}=1$, $a=1,2$.
We shall denote by $\a_1$ and $\a_2$ the simple roots and the highest is
$\psi = \a_1 + \a_2$ which we normalize as $\psi^2=2$. We
also have $K_{\psi 1}=K_{\psi 2}=1$.
The structure constants $\eps (\a , \b )$, appearing in \rf{km},
can be chosen such that  $\eps (\a , \b ) = -\eps (-\a , -\b )$,
$\eps (\a_1 , \a_2 ) = 1$, $\eps (-\a_1 , \psi ) = 1$ and
$\eps (-\a_2 , \psi ) = -1$.
The basis discussed in section \ref{sec:cat} is given by
\br
E_{3n+1} &=& E_1^n + E_2^n + E_{-3}^{n+1} \nonu\\
E_{3n-1} &=& E_{-1}^n + E_{-2}^n + E_{3}^{n-1} \nonu\\
F^1_{3n+1} &=& E_1^n + \g E_2^n + \g^2 E_{-3}^{n+1} \nonu\\
F^1_{3n} &=& H_1^n + \g H_2^n + \g^2 H_{0}^{n} \nonu\\
F^1_{3n-1} &=& E_{-1}^n + \g E_{-2}^n + \g^2 E_{3}^{n-1} \nonu\\
F^2_{3n+1} &=& E_1^n + \g^2 E_2^n + \g E_{-3}^{n+1} \nonu\\
F^2_{3n} &=& H_1^n + \g^2 H_2^n + \g H_{0}^{n} \nonu\\
F^2_{3n-1} &=& E_{-1}^n + \g^2 E_{-2}^n + \g E_{3}^{n-1}
\lab{sl3basis}
\er
where $H_{0}^{n} = -H_1^n - H_2^n + \d_{n,0} C$, and where [Ball generators
are defined in the Chevalley basis and satisfy the
commutation relations \rf{km}.
The parameter $\g$ is a cubic root of unity, i.e. $\g^3 =1$.

Using the procedure of section \ref{sec:charges} one can check that the
first two right charge densities are given by
\br
\ca_{+}^{R,(1)} &=& {1\o 3} \( (\pa_{+} \vp_{1})^2 +
(\pa_{+} \vp_{2})^2 -
\pa_{+} \vp_{1} \pa_{+} \vp_{2} + \pa_{+} \eta \pa_{+} \vp_{1} +
\pa_{+} \eta \pa_{+} \vp_{2} \right.  \nonu\\
&+& \left. 3 \pa_{+} \eta \pa_{+} \nu -  \pa_{+}^2 \vp_{1} -
\pa_{+}^2 \vp_{2} - 3 \pa_{+}^2 \nu \) \nonu\\
\ca_{-}^{R,(1)} &=& 0
\er
and
\br
a_{+}^{R,(2)} &=& {1\o 3} e^{-2\int^{x_{+}} J^R} \( (\pa_{+} \vp_{1})^2
\pa_{+} \vp_{2} -
(\pa_{+} \vp_{2})^2 \pa_{+} \vp_{1} + {1\o 2}
\pa_{+} \vp_{1} \pa_{+}^2 \vp_{2} -  {1\o 2}  \pa_{+}
\vp_{2} \pa_{+}^2 \vp_{1} \) \nonu\\
a_{-}^{R,(2)} &=& {1\o 6} e^{-2\int^{x_{+}} J^R} \( q_1\, \pa_{+}
\vp_{2} e^{2 \vp_{1} - \vp_{2} + \eta} - q_2 \, \pa_{+}
\vp_{1} e^{2 \vp_{2} - \vp_{1} + \eta} \right. \nonu \\
&+& \left. q_0 \, (\pa_{+} \vp_{1} - \pa_{+} \vp_{2})
e^{- \vp_{1} - \vp_{2} + \eta} \)
\er

\sect{Discussion}
\label{sec:disc}
We constructed an infinite number of conserved charges for the Conformal
Affine Toda models. These charges are intrinsically related to the Heisenberg
subalgebra $\hlie$ of the underlying KM algebra. As we have pointed out an
interesting aspect of those conservation laws is that, although the CAT model
is a conformally invariant model, its symmetries are not described by chiral
currents only. The chiral charges we constructed do not possess chiral
densities, except of course for the stress tensor components and the spin $1$
currents $\pa_{\pm}\eta$. Therefore not all the symmetries give rise to a
$W$-algebra \ct{hspin,deform,2boson}.

In ref. \ct{hspin} we have shown how to obtain the chiral densities of the CAT
model via the Hamiltonian reduction as the remaining KM currents of the
two-loop WZNW model \ct{AFGZ}. The densities were obtained, for each chiral
sector, by constraining the KM currents and then by choosing appropriate gauge
fixing conditions such that the so called remaining currents depend on the CAT
model fields only. However, we have not succeeded  in obtaining the whole
spectrum of CAT model charges via Hamiltonian reduction. We now comment that
this is possible at the level of the conservation laws by considering the
constraints on both chiral sectors.
Let us consider for instance the right currents. After the constraints are
imposed they are given by \ct{hspin}
\br
J_{red} &=& k ( {\hat g}^{-1} \pa_{+} {\hat g} )_{red} =
{\cal M}^{-1} ( \pa_{+} \Phi + E_1 ) {\cal M} + {\cal M}^{-1} \pa_{+} {\cal M}
\nonu \\
&=& E_1 + \sum_{M>0} j_M E_{-M} + \sum_{m>0} \sum_{a=1}^r j^a_m F^a_{-m}
\lab{redcur}
\er
where ${\cal M}$ is an exponentiation of the negative step operators of the KM
algebra and appear in the Gauss decomposition ${\hat g}= {\cal N} A {\cal M}$
\ct{hspin}. Notice that  \rf{redcur} looks very similar to \rf{gtaplus}.
However the quantities entering $g_R$ in \rf{gtaplus} are arbitrary parameters
whilst those in ${\cal M}$ are WZNW fields. In \ct{hspin} we have shown that
one has to gauge fix the currents $j^a_m$ and $j_M$, for $M>1$ in order to
eliminate the unwanted WZNW fields in ${\cal M}$. We were left then with only
two remaining currents.  If one does not gauge fix all the $j_M$ currents, they
will be remaining chiral currents depending on some unwanted fields. However,
using the constraints on the left sector one can eliminate such fields when
those currents are inserted into the equations describing the conservation
laws.  Such mixture of constraints on both chiral sectors accounts for the non
chiral character of the charge densities.
Therefore, the Hamiltonian reduction provides the
same conservation laws we obtained in this paper, although it does not give
directly the expression for the charge densities.
We now make some comments on the relation among the conserved charges for the
different hierarchies of Toda models. One notices that by performing global
gauge transformations of the type \rf{global} one can eliminate the coupling
constants from one of the components of the gauge potential. In particular, we
have chosen the gauge such that the charges \rf{chargesR} and \rf{chargesL} are
independent of the coupling constants. By taking the limit $q_0 \rightarrow 0$
one observes from \rf{todathree} that the $\nu$ field becomes free
and the second exponential on the r.h.s. of \rf{todaone} is dropped. Since the
charges \rf{chargesR} and \rf{chargesL} are unchanged under such limit they are
also conserved charges of the model obtained by that limit. In particular, by
taking the special case where $\eta = 0$, one observes that the models under
consideration are the Affine Toda ($q_0 \neq 0$) and Conformal Toda ($q_0 =0$).
Of course, they both carry the extra passive field $\nu$, which does not affect
the $\vp$'s dynamics.  Let us remark however that although the charge densities
$a_{+}^{R,(M)}$ depend explicitly on the field $\nu$, the densities constructed
in appendix A do not.  Therefore, since the field $\nu$ can really be droped
from the integrated charges, we have shown that the chiral charges of the AT an
CT models are the same. In the case of the CT model we already know that not
only the charges but also the densities are chiral. This  then generalize the
result already known in the literature about the relation between the AT and CT
charges for the case of $Sl(N)$ \ct{chodos,west,ff}.

Finally, we would like to mention that the corresponding FPR \rf{fpr} for the
chiral components of the gauge potentials contains $\d^{\pr}$ terms.  The
involution of the chiral charges therefore is not as clear as that for the
charges conserved in time. It would be interesting to verify whether all the
light-cone charges share that property.

\appendix
\section{Appendix: $\nu$ independence of the charges}
\lab{appendixa}
\setcounter{equation}{0}

We now show that it is possible to construct another set of chiral charges,
obviously related to the ones given above, which are independent of the field
$\nu$. The construction works similarly for both chiralities, but here we will
discuss only the right charges. The basic idea is to perform a gauge
transformation to eliminate the field $\nu$ from the $A_{+}^R$ component of the
gauge potentials given in \rf{rp}. One point about this procedure is that we
will not get a chiral density, like we did before, corresponding to one of the
components of the CAT model energy-momentum tensor.

It is useful to work with the basis of the subspace $\flie_0$. As we pointed
out below \rf{e1}, all subspaces $\flie_n$ have dimension equal to the rank of
$\lie$ ($\equiv r$), except for $\flie_0$ which has dimension $r+2$. Let us
denote by $\flie_0^a$ ($a=1,2,\ldots r$), $C$ and $\rh$ (defined in \rf{rho})
the basis of $\flie_0$. The generators  $\flie_0^a$ are linear combinations of
$H_a^0$ and $C$. Therefore we can write $\Phi$ defined in \rf{Phi} as
\be
\Phi = \sum_{a=1}^{r} \vt^a \flie^a_0 + \eta {\hat \rho} +
\sigma C
\lab{Phi2}
\ee
where
\be
\vt^a \equiv \mbox{\rm linear combination of $\varphi$'s (no $\nu$)}
\ee
and
\be
\sigma \equiv \nu + \mbox{\rm linear combination of $\varphi$'s}
\ee
We now perform a gauge transformation like \rf{gt} on $A_{\pm}^R$ with $g=
e^{-\sigma C}$ and get
\br
A^R_{+} &\rightarrow & \at_{+} = \pa_{+} \pt +  E_1 \nonu \\ A^R_{-}
&\rightarrow & \at_{-} = e^{-\pt} {\cal E}_{-} e^{\pt} - \pa_{-} \sigma C
\lab{rp2}
\er
where
\br
\pt &=& \sum_{a=1}^{r} \vt^a \flie^a_0 + \eta {\hat \rho}
\er
The only dependence on the $\nu$ field, or equivalently $\sigma$, is on the
$\at_{-}$ component of the gauge potential.  Notice we have replaced $\Phi$ by
$\pt$ on the exponential term in the second equation in \rf{rp2}, because the
$C$ term is irrelevant in the conjugation.

We can now rotate the gauge potentials $\at_{\pm}$ into the Heisenberg
subalgebra by performing a gauge transformation like \rf{gtaplus} but with
$g_R$ replaced by ${\tilde g}_R$ and
\be
{\tilde g}_R  \equiv e^{(\sum_{m>0} \zeta_{-m})} \, \, \, , \, \, \,
\mbox{\rm where} \, \, \zeta_{-m} \in \flie_{-m}
\ee
i.e., ${\tilde g}_R$ does not have the term in the direction of $E_{-1}$.

Denoting the transformed $\at_{\pm}$ by $\atc_{\pm}$ and writing $\atc_{+} =
\sum_{m=-1}^{\infty} \atcm{-m}_{+}$, we obtain that $ \atcm{-m}_{+}$ are given
by the expressions \rf{expansion} with $\Phi$ replaced by $\pt$ and by droping
the $\chi$ terms. One notices that $\atcm{0}_{+}$ can have components in the
direction of $\flie^a_0$ and ${\hat \rho}$ but not in the direction of $C$. So
we choose the $r$ parameters of $\zeta_{-1}$ to kill the $r$ components of
$\atcm{0}_{+}$ in the direction of $\flie^a_0$. Since the ${\hat \rho}$ is not
affected by the term $\lb E_1 \, , \, \zeta_{-1}\rb $, we get $\atcm{0}_{+} =
\pa_{+} \eta {\hat \rho}$. From now on we can use the $r$ parameters of
$\zeta_{-m-1}$ to kill the $r$ components of $ \atcm{-m}_{+}$ in the direction
of $\flie^a_{-m}$. So we have shown that
\be
\atc_{+} = E_1 + \pa_{+} \eta {\hat \rho} + \sum_{M>0} \atcm{M}_{+} E_{-M}
\ee
Notice that $\atc_{+}$ is $\sigma$ (and so $\nu$) independent.

Using arguments similar to those used in \rf{gtaminus}-\rf{b1} we obtain that
\be
\atc_{-} = \sum_{M>0} \atcm{M}_{-} E_{-M} - \pa_{-} \sigma C
\ee
Notice however that $\atcm{1}_{-} \neq 0$, and so $\atcm{1}_{+}$ is not chiral.

{}From the zero curvature for these potentials we get
\be
\pa_{+} \pa_{-} \sigma = {1\o h} \mbox{\rm Tr} \( E_1 E_{-1}\) \atcm{1}_{-}
\ee
and
\be
(\pa_{+} - M J^R ) \atcm{M}_{-} - \pa_{-} \atcm{M}_{+} =0 \, \,
\, , \, \,  \, \mbox{\rm for $M>0$}
\lab{a10}
\ee
with $J^R = \pa_{+} \eta$.

We write \rf{a10} as
\be
\pa_{+} {\tilde a}_{-}^{R,(M)} - \pa_{-} {\tilde a}_{+}^{R,(M)} =0 \, \,
\, , \, \,  \, \mbox{\rm for $M>0$}
\ee
where
\be
{\tilde a}_{\pm}^{R,(M)} \equiv e^{-M \eta_{+}} \atcm{M}_{\pm}\, \,
\, , \, \,  \, \mbox{\rm for $M>0$}
\ee
and where we have used the fact that $\eta (x_{+},x_{-}) = \eta_{+}(x_{+}) +
\eta_{-}(x_{-})$.

Therefore the charges
\be
\int dx_{+} {\tilde a}_{+}^{R,(M)}\, \,
\, , \, \,  \, \mbox{\rm for $M>0$}
\ee
are chiral and independent of the field $\nu$ (or equivalently $\sigma$).

\lskip
{\bf Acknowledgements}
We gratefully acknowledge support within CNPq/NSF Cooperative Science Program.
One of us (HA) thanks Instituto de F\'{\i}sica Te\'{o}rica-UNESP
for kind hospitality.

\end{document}